\documentclass{iopart}

\usepackage{graphicx}

\usepackage{wasysym}

\def\cof{coefficients for Lorentz violation}

\def\Kost{Kosteleck\'y V A}
\def\prd#1{{\it \PR D}\ {\bf #1}}
\def\prl#1{{\it \PRL}\ {\bf#1}}

\def\al{\alpha}
\def\be{\beta}
\def\ga{\gamma}
\def\de{\delta}
\def\ep{\epsilon}

\def\et{\eta}

\def\ka{\kappa}
\def\la{\lambda}

\def\si{\sigma}

\def\ch{\chi}
\def\ps{\psi}

\def\Ga{\Gamma}

\def\La{\Lambda}

\def\cl{{\cal L}}
\def\cL{{\cal L}}

\def\mn{{\mu\nu}}

\def\fr#1#2{{{#1} \over {#2}}}
\def\half{{\textstyle{1\over 2}}}

\def\frac#1#2{{\textstyle{{#1}\over {#2}}}}

\def\lsim{\mathrel{\rlap{\lower4pt\hbox{\hskip1pt$\sim$}}
    \raise1pt\hbox{$<$}}}
\def\gsim{\mathrel{\rlap{\lower4pt\hbox{\hskip1pt$\sim$}}
    \raise1pt\hbox{$>$}}}

\def\prt{\partial}

\def\etal{{\it et al.}}

\def\pt#1{\phantom{#1}}

\def\ol#1{\overline{#1}}

\def\vb#1#2{e_{#1}^{{\pt{#1}}#2}}
\def\ivb#1#2{e^{#1}_{{\pt{#1}}#2}}
\def\uvb#1#2{e^{#1#2}}

\def\acc{{\rm a}}

\def\lrpartial{\raise 1pt\hbox{$\stackrel\leftrightarrow\partial$}}

\def\lrprtmu{\stackrel{\leftrightarrow}{\partial_\mu}}

\def\lrDmu{\stackrel{\leftrightarrow}{D_\mu}}
\def\lrDnu{\stackrel{\leftrightarrow}{D^\nu}}
\def\lrvec#1{\stackrel{\leftrightarrow}{#1} }

\def\a{$a_\mu$}
\def\b{$b_\mu$}
\def\c{$c_{\mu\nu}$}
\def\d{$d_{\mu\nu}$}
\def\e{$e_\mu$}
\def\f{$f_\mu$}
\def\g{$g_{\la\mu\nu}$}
\def\H{$H_{\mu\nu}$}

\def\kaf{\hat k_{AF}}
\def\kafd#1{k_{AF}^{(#1)}}

\newcommand{\beq}{\begin{equation}}
\newcommand{\eeq}{\end{equation}}
\newcommand{\bea}{\begin{eqnarray}}
\newcommand{\eea}{\end{eqnarray}}
\newcommand{\bit}{\begin{itemize}}
\newcommand{\eit}{\end{itemize}}
\newcommand{\rf}[1]{(\ref{#1})}

\begin{document}

\article{KEY ISSUES}{What Do We Know About Lorentz Invariance?} 


\author{Jay D.\ Tasson}

\address{
Physics and Astronomy Department,
Carleton College,
Northfield, MN 55901, USA}
\ead{jtasson@carleton.edu}
\begin{abstract}
The realization that Planck-scale physics can be tested
with existing technology through the search for spacetime-symmetry violation
brought about the development of a comprehensive framework,
known as the gravitational Standard-Model Extension (SME),
for studying deviations from exact Lorentz and CPT symmetry in nature.
The development of this framework and its motivation
led to an explosion of new tests of Lorentz symmetry over the past decade
and to considerable theoretical interest in the subject.
This work reviews the key concepts associated with Lorentz and CPT symmetry,
the structure of the SME framework,
and some recent experimental and theoretical results.

\end{abstract}

\maketitle

\section{Introduction}

Lorentz symmetry is a foundational assumption of 
both of our current best theories of physics: 
the Standard Model of particle physics
and General Relativity.
At the heart of Lorentz symmetry is the Principle of Relativity:
the property of nature
that experimental results do not seem to depend 
on the orientation of the laboratory
(rotation invariance)
or its velocity though space (boost invariance).
The principle of relativity is a very old idea.
In the context of mechanics,
it can be traced back to at least the time of 
Galileo Galilei \cite{Gali1632}.
The Principle of Relativity requires
that the laws of physics take the same form,
independent of the velocity 
and orientation of the experiment.

During the years leading up to 1905,
the Principle of Relativity received a great deal of attention.
The mathematical transformation,
now known as the Galilean transformation,
which was thought to implement the change from one velocity to another
did not appear to apply to the theory of electromagnetism
as newly unified by Maxwell.
A new transformation,
the Lorentz transformation,
found partly by Lorentz \cite{Lore},
was given its current interpretation by Einstein \cite{Eins1905}.
Einstein's interpretation forever changed our understanding of 
time, as well as other physical quantities such as energy,
and in the process asserted that the Principle of Relativity
applies to all phenomena.
Lorentz symmetry is a global symmetry
of the Standard Model in flat spacetime,
and a local symmetry among freely-falling frames
in General Relativity.

Though tests of Lorentz symmetry 
have been performed since the time of Einstein,
the past several decades
have seen 
considerable interest in the subject \cite{KostProc}
and an explosion of new tests \cite{data}.
These tests have been performed across a wide range
of fields of physics
and have in some cases reached impressive sensitivity \cite{data},
though many ways in which Lorentz symmetry could be violated remain untested.
The primary motivation for this resurgence of interest
is the search for new  physics at the Planck scale \cite{KostSamu1989},
though placing known physics on a sound experimental foundation
also offers a motivation.

In the 1990s,
a comprehensive effective field-theory based framework 
for studying Lorentz symmetry known as the
gravitational Standard-Model Extension (SME) was developed \cite{CollKost1998}.
The SME contains the Standard Model and General Relativity
as well as all possible Lorentz-violating terms
that can be constructed from the associated fields.
Most of the recent high-sensitivity tests
have been motivated and analyzed in this framework,
and much theoretical work
has also been done in this context.
Hence the SME is the primary framework
for discussion in this review.
It should be noted that since the time of Einstein,
a great deal of work has been done on the topic of Lorentz symmetry,
far more than can be considered in this short review.
The goal of this work is to review some key aspects
and recent developments
related to the effective field-theory approach
to the search for Lorentz violation.

This paper is organized as follows.
Section \ref{sec:basics}
presents some foundational material
including a discussion of the role of symmetry in fundamental physics,
comments on the relation of Lorentz symmetry to other symmetries,
a simple example of rotation-invariance violation,
and some additional discussion of the motivation 
for considering symmetry violation.
The rational for studying the topic of Lorentz symmetry
using the effective field-theory based framework
provided by the SME is considered in section \ref{sec:sme}.
Section \ref{sec:structure}
addresses the construction of the SME,
presents some popular limiting forms,
and reviews some theoretical studies of its structure.
In section \ref{sec:tests},
a discussion of some key considerations associated with experimental
and observational work on Lorentz symmetry in the context of the SME
is provided along with some discussion of the modern high-sensitivity tests
that have been performed
and the remaining space open to investigation.
Motivated by current events
and the possibility of explaining observed physics beyond the Standard Model,
this section also provides some additional comments
on neutrinos.
Finally section \ref{conclusion} provides a summary
and some conclusions regarding the current state of the field
of Lorentz-symmetry study.

\section{Basics}
\label{sec:basics}

This section provides some basic information
and simple examples useful in understanding
the development to follow.
We first consider symmetry in physics generically
and the relation of Lorentz symmetry to other symmetries
in section \ref{symmetry}.
We then consider rotation invariance
as a simple example
of both symmetry
and of symmetry violation
in section \ref{rotation}.
Section \ref{motivation}
expands on the motivation for considering
the violation of spacetime symmetries.

\subsection{Symmetry}
\label{symmetry}

Symmetry in physics can be informally stated as a transformation on a system
that leaves it unchanged.
This approximately matches the everyday use of the word
in which one might say that a sphere exhibits rotational symmetry
since it looks the same when rotated.
A physical symmetry
then corresponds to a mathematical operation
on the laws that describe the system
such that there is no effect (up to coordinate choices)
on the observable quantities associated with the system
predicted by those laws.
Symmetry has long been a guiding principle
in physics,
and in modern theoretical physics,
it is given a lead role \cite{symmetry}.
In proposing a theory,
the desired symmetries are usually one of the foundational assumptions.
A lagrangian is then written
to contain all possible terms consistent with the symmetries
and particles that have been assumed.

Lorentz invariance is an example
of a symmetry in physics,
which contains
two subgroups:
rotations and boosts.
It is a spacetime symmetry since 
it is associated with transformations
in the physical space.
Other symmetries
closely related to Lorentz symmetry
include diffeomorphism invariance
and the discrete symmetries,
which include charge conjugation C, parity P, time reversal T,
and combinations of these.

Of particular interest in the present context
is the combination of all three discrete symmetries CPT
due to its close connection with Lorentz symmetry
as well as the fact that it is the only discrete symmetry
or combination of discrete symmetries
that is not violated in the Standard Model.
There is a well-known result
called the CPT theorem,
which roughly states that local quantum field theories with Lorentz symmetry,
including those used to describe known particle physics,
also have CPT symmetry.
For additional discussion of this result,
see \cite{CPT} and references therein.
In 2002 a particularly strong form of the CPT theorem 
was proven by Greenberg \cite{Gree2002}:
``An interacting theory that violates CPT invariance necessarily violates
Lorentz invariance.'' 
This implies that CPT violation consistent
with the framework of conventional quantum field theory
of known physics is described 
by the SME. 

Note that throughout this work,
references to Lorentz violation
refer generically to the possibility of arbitrary violations
of rotation or boost invariance.
In particular,
the existence of a preferred frame in which physics is isotropic
is not assumed.

\subsection{Rotation}
\label{rotation}

Here we consider rotation symmetry,
a part of Lorentz invariance,
as a concrete
example of symmetry.
For simplicity,
we consider 
the role of rotation symmetry 
in classical mechanics.
We first consider a system with full rotation invariance.
We then consider effective rotation-invariance violation
as well as more fundamental rotation-invariance violation.

As an example of rotation invariance,
consider a particle of mass $m$
under the influence of a quadratic potential:
\beq
L = \half m \dot{\bi{r}}^2 - \half k \bi{r}^2,
\eeq
where $\bi{r}$ is a position vector in 3 dimensions and
$k$ is a constant.
Such a system could be constructed in a lab
by attaching one end of a spring to a fixed point in a box (call it the origin)
and attaching the other end to the particle.
A rotation of this system can be carried out by instructing a worker in the lab
to rotate the box.
Mathematically this rotation can be carried out by applying an appropriate rotation matrix 
to the position vector, $r_{j^\prime} = R_{j^\prime k} r_k$.
Explicit calculation will reveal that the lagrangian is unchanged
and hence scientists in the lab
will find identical observables associated with the rotated system
for a given set of initial conditions.

In general,
there are two types of Lorentz transformations that can by applied to a system:
{\it Observer Transformations} and {\it Particle Transformations}.
In the above example,
we have performed a particle transformation,
that is,
we transformed,
rotated in this case,
the particles and fields involved in the experiment by instructing
workers in the lab to physically rotate the experiment.
An observer transformation in this case
would involve choosing new coordinates.
In the above example,
which has rotation invariance,
these two transformations can be chosen to have the same effect.
This fact is a direct consequence of rotation invariance.

Next consider an example that has effective rotation-invariance violation.
Consider a particle in a linear potential,
an approximate model for a particle in an Earth-based laboratory,
for example.
Such a system may be mathematically described via the lagrangian,
\beq
L = \half m \dot{\bi{r}}^2 - m \bi{g} \cdot \bi{r},
\eeq
where $\bi{g}$ is a constant background gravitational field
and $\bi{r}$ is a position vector measured from some origin
within some box in the lab.
A rotation of the experiment in the lab,
a particle rotation,
could again be carried out by having a worker rotate the box,
and such a transformation would be carried out mathematically
by applying the rotation matrix to the position vector
for the particle
as before $r_{j^\prime} = R_{j^\prime k} r_k$.
Since $\bi g$ is unaffected by this rotation of the box
in the lab,
the rotation matrix should not be applied to this vector.
The lagrangian will not be invariant under such a transformation.
The transformed lagrangian will take the form
\beq
L = \half m \dot{\bi{r^\prime}}^2 - g_k R_{k j^\prime} r_{j^\prime}.
\eeq
This will have observable consequences.
For example,
in the unrotated system,
the particle may fall toward a face
of the box,
where as in the rotated system
it may fall toward a corner
or a different face of the box.

I refer to the type of symmetry violation
in the above example 
as ``effective'' because it is not a consequence
of rotation-invariance violation
in the fundamental laws of physics.
It is instead a consequence of the fact
that a true ``rotation of the experiment''
requires turning the Earth as well,
since it is a part of this experiment.
If we did so,
the rotation transformation would be applied to $\bi{g}$
as well,
and we would see that the system
is particle-rotation invariant.
This idea of effective Lorentz violation
has been applied to investigations of gravitomagnetism \cite{jt2012}
and has been used to obtain some of the best constraints
on spacetime torsion \cite{KostRuss2008}.

Note that an observer transformation
could also be applied in the above example.
This would involve rotating our coordinates in the lab.
Under such a transformation,
the rotation matrix would be applied to all vectors.
The workers would make no adjustments to the physical system,
and the particle would still fall toward the same face of the box.
The only change would be in the name given to that direction.
If the workers had originally called it the $\hat x$ direction,
they might now call it the $\hat y$ direction.

As a final example,
let us suppose that there is a hypothetical particle in our universe
that experiences an acceleration of magnitude $\acc = F/m$,
when a force of magnitude $F$ is applied in a given direction.
Suppose further that the same particle experiences an acceleration of magnitude
$\acc^\prime = F/m^\prime$
when the force of magnitude $F$ is applied in an orthogonal direction \cite{Bert}.
Assume
that such an effect is found to be equally valid independent of the type of force,
location in the universe, or other such factors.
Such an effect would violate rotation invariance,
but not due to the fact that some aspect of the experiment has been neglected
as in the example of the gravitational field.
It is an effect that could be attributed to the spacetime vacuum itself.
A complete nonrelativistic theory having this feature can be written,
\beq
L = \half m_{jk} \dot r_j \dot r_k - U(\bi{r}).
\eeq
Here $m_{jk}$ does not transform under particle transformations,
but is a 2-tensor under observer rotations.
This model can arise as a limit of the SME \cite{Bert}.
Here we have illustrated the basic idea of spacetime-symmetry violation
using rotations due to the simple visual nature of this limit,
however the same ideas apply to boost invariance
and other spacetime symmetries.

\subsection{Motivation}
\label{motivation}

There are two common motivations
for searching for Lorentz violation.
Perhaps the most exciting is the possibility
of detecting new physics at Planck scale
with existing technology.
Another motivation is provided
simply by the desire to put a foundational principle
of both of our current best theories,
the Standard Model and General Relativity,
on the strongest experimental grounds possible,
since history demonstrates that deeply held principles
are worth rigorous testing.

Though the Standard Model and General Relativity
form an impressive description of physics at presently accessible energy scales,
it is believed that they are merely the low-energy limit
of a single quantum-constant theory
at the Planck scale, $10^{19}$ GeV.
Obtaining experimental information 
to guide the development of a theory of Planck-scale physics
is notoriously difficult.
For example,
it is difficult to imagine probing this energy directly via
particle-accelerator experiments
as the Planck scale remains more than 15 orders of magnitude
beyond the energy of the LHC.

An alternative approach is to search for Planck-suppressed deviations
from known physics.
The idea is to search for Planck-scale physics
using Planck sensitivity rather than Planck energy.
In fact,
the realization that natural mechanisms for 
the generation of Lorentz violation
exist in string theory \cite{KostSamu1989,motivation:string,KostPott1995},
a leading candidate for the underlying theory,
triggered much of the recent interest in Lorentz violation.
Since that time,
other scenarios for underlying theory compatible
with Lorentz violation have been found including ones based on
noncommutative field theories
\cite{honc,motivation:ncfield}, 
spacetime-varying fields \cite{motivation:svf}, 
quantum gravity \cite{motivation:qg,MyerPosp2003}, 
random-dynamics models \cite{motivation:rdm}, 
multiverses \cite{motivation:multiverse}, and
brane-world scenarios \cite{motivation:bws}.

Returning to the level of known physics,
Lorentz symmetry is a foundational principle
of both the Standard Model and General Relativity.
This fact implies that the observation of Lorentz violation
would be a strong indicator of new physics.
The foundational nature of the idea also suggests
that it should be placed on a strong experimental foundation.
The history of physics suggests that broad testing of foundational assumptions,
particularly those based on the beauty of symmetry,
is well worthwhile.

The beauty of perfect symmetry
is perhaps among the oldest ideas in physics
and has perhaps even been given additional importance over time.
Such ideas predate even modern science itself
in Aristotle's assertion of perfectly-circular planetary orbits.
As already noted,
the Principle of Relativity in mechanics,
as articulated by Galileo,
was based on rotation invariance
and Galilean-boost invariance.
In modern theoretical physics,
symmetries including Lorentz symmetry,
the discrete symmetries,
and gauge invariance
are central.
Although the beauty of perfect symmetry is appealing,
historical examples of symmetry breaking are also
prevalent.
We are well aware that the planetary orbits are not perfect circles,
every combination of the discrete symmetries
is known to be broken except CPT,
and electroweak-symmetry breaking is a key feature of the Standard Model.

A broad search appears to be a reasonable approach
in searching for symmetry violation.
For example,
no test of parity symmetry with electromagnetic experiments,
no-matter how sensitive,
can exclude the possibility of parity violation in nature
as parity violation is observed in weak-interaction physics
but not electromagnetic physics.
Such examples serve as motivation
for a broad search for Lorentz violation.

\section{A comprehensive theory: the SME}
\label{sec:sme}

As noted in the examples
in section \ref{symmetry},
the classic approach to testing Lorentz symmetry
is to do an experiment 
having a given orientation and boost,
then repeat the experiment with a different orientation or boost
and attempt to detect a difference.
In principle this can be done randomly
with any experiment.
However proceeding in a random way
has key disadvantages.
First,
there is no way of making quantitative comparisons
between very different types of experiments.
This makes it hard to answer questions
about whether or not it is worth devoting resources
to improve atomic-clock tests
or whether one should do accelerator experiments instead.
Second,
it is often unclear which aspects of the experiment
are most relevant to likely theoretical possibilities.
For example,
early CPT tests often averaged data taken over a period of time;
however,
since CPT violation comes with Lorentz violation,
this averaging typically hides periodicity due to changes in orientation
of the experiment.
Hence proceeding without a comprehensive theory
can lead to choices that mask a key signal.
Finally,
each theoretical model must confront experiment
independently,
with no easy way of cataloguing known constraints.

The above issues are addressed by the SME,
which is a comprehensive field-theory based test framework
in which to explore Lorentz violation.
The framework consists of known physics
in the form of the Standard Model plus General Relativity
action
along with all possible Lorentz-violating terms
that can be constructed from the associated fields.
This has been done explicitly
for the case of power-counting renormalizable mass dimension 3 and 4 operators
in each sector of the Standard Model \cite{CollKost1998}
as well as for the lowest-order terms in the gravity sector 
under the assumption of Riemann-Cartan geometry \cite{Kost2004}.
Higher mass-dimension operators have also been classified
in several sectors \cite{KostMewe2009,KostMewe2012,KostMewe2013}.
Standard lore holds that the Standard Model and General Relativity
are the low-energy limit of a unified theory
at the Planck scale.
Assuming a smooth match between physics at our current energies
and Planck-scale physics,
the Standard Model and General Relativity
could be regarded as leading terms in a series approximation
of the unified theory,
with terms involving operators of higher mass dimension
viewed as high-order corrections.
The SME action
provides the complete series
for Lorentz-violating physics.

The key advantage of the SME
is that it offers a complete field theory
incorporating all possible forms of Lorentz and CPT violation
that can be constructed from the associated fields.
As a complete effective field theory,
the SME provides complete predictive power.
The outcome of any experiment can in principle
be calculated.
This allows one to predict {\it a priori} which experiments
will attain the best sensitivity.
It is possible to know which types of experiments measure the same thing
and which measure different things,
and such conclusions will be true in any consistent field-theoretic model.
Thus it provides a framework for cataloguing results,
forming a very definite list of constraints
against which individual models can be compared.
As a complete theory incorporating Lorentz violation,
the SME is also very useful 
in considering theoretical aspects of Lorentz violation.

It is important to emphasize 
that the SME is not a model,
rather it is a test framework designed for a broad search.
Since the Standard Model and General Relativity
have passed all experimental challenges thus far,
the plan is to do a broad and comprehensive search for new physics
rather than build specific models.
The idea is that
a more effective time for model building
would be after 
a result is observed that is not consistent with 
the Standard Model and General Relativity,
thus providing likely new physics to be modeled.

It should also be noted that a variety of other 
specialized formalisms exist
that consider Lorentz violation.
For a review of various approaches,
see  \cite{otherframework}.
Examples that are consistent with effective field theory
are contained in the SME framework.
Section IV F of  \cite{KostMewe2009}
and Section VI B of \cite{KostMewe2013}
demonstrate how several examples
can be recast in the language of the SME.
A common choice in specialized formalisms,
including some constructed as special limits of the SME,
is to consider models
in which there exists an observer frame where rotation invariance is preserved, and Lorentz violation is associated purely with boost violation.
This frame is often identified with the rest frame
of the cosmic microwave background radiation,
though other choices have also been considered.
Such models are sometimes referred to as isotropic models,
or with tongue-in-cheek as `fried-chicken models' \cite{KostMewe2004}
in the literature.
The latter name was coined by analogy with a popular food 
in the United States due to the popularity and simplicity of the models.
The idea being that fried chicken is good,
and everyone likes it,
but if that is all one eats,
one misses a lot.
The frame in which physics is isotropic in such models is usually referred to
as a preferred frame
and associated effects are often referred to as preferred-frame effects.
The term preferred-frame effects
is sometimes used synonymously with Lorentz violation;
however,
this is not correct.
In the SME the existence of a preferred frame
is not assumed
and generically the existence of Lorentz violation
may not generate such a preferred frame.
It should also be emphasized that the term isotropic
is misleading,
since even in isotropic models,
physics is isotropic in only one frame.
In all other frames,
rotation invariance is violated.
Moreover,
the isotropic frame cannot be chosen 
as the frame of an Earth-based laboratory,
and rotation-invariance violation will be present in Earth-based experiments
even in isotropic models.

\section{Structure of the SME}
\label{sec:structure}

The Lorentz-violating terms of the SME
are constructed by coupling observer vector or tensor \cof\
to Standard-Model operators.
As an example,
consider the following Lorentz-violating terms occurring in the quark sector:
\bea
\cl^{\rm CPT-even}_{\rm quark} &=& 
\half i (c_Q)_{\mu\nu AB} \overline{Q}_A \ga^{\mu} \lrDnu Q_B
\nonumber\\ &&
+ \half i (c_U)_{\mu\nu AB} \overline{U}_A \ga^{\mu} \lrDnu U_B 
\nonumber\\ &&
+ \half i (c_D)_{\mu\nu AB} \overline{D}_A \ga^{\mu} \lrDnu D_B.
\quad
\label{lorviolq}
\eea
Here capital Latin indicies are flavor indicies,
while Greek indicies are spacetime indicies.
The Standard Model covariant derivative is denoted $D_\mu$,
and the operation of the derivative on arbitrary objects $A,B$
is defined as $A\lrprtmu B \equiv A\prt_\mu B - (\prt_\mu A) B$.
The notation
\beq
Q_A = \left( \begin{array}{c} u_A \\ 
d_A \end{array} \right)_L ~~ , ~~ 
U_A = (u_A)_R ~~ , ~~ D_A = (d_A)_R
\quad , 
\eeq
is used for the left- and right-handed 
quark multiplets, where
$A = 1,2,3$ labels the flavor.
The objects 
$(c_Q)_{\mu\nu AB}$,
$(c_U)_{\mu\nu AB}$,
and $(c_D)_{\mu\nu AB}$ are examples of \cof.
The size of these coefficients controls the amount of Lorentz violation
of the given type in the theory.
Note that each type of particle
has its own \cof,
reflected here by the existence
of the 3 coefficients shown explicitly
each having flavor indices.
Thus the generality of the SME allows
for the possibility that Lorentz violation
could exist in nature associated with one particle
but not another.
Such generality is important
since history indicates that finding new physics
requires looking in the right place
as noted in section \ref{motivation}.
The notation
$\cl^{\rm CPT-even}_{\rm quark}$
for the partial Lagrange density
shown here
indicates that it is the part of the lagrangian
associated with the quarks
containing \cof\
that do not violate CPT symmetry.
Other partial Lagrange densities
exist in the theory that contain CPT-violating
\cof.
The CPT even terms are easily recognized as they
involve coefficients with an even number of spacetime indices,
while CPT odd terms involve an odd number of spacetime indices.

The basic framework for including gravity
in the general action-based construction of the Lorentz-violating
SME was developed in 2004 \cite{Kost2004}.
Riemann-Cartan spacetimes,
which allow for nonzero torsion in addition
to curvature, were considered.
Such spacetimes can be reduced to the Riemann spacetime
of General Relativity in the appropriate limit.
The systematic incorporation of Lorentz-violating operators
in the gravitational-sector action was developed,
and gravitational couplings were introduced in the other sectors of the theory.
The consideration of Lorentz violation in nontrivial geometries 
led to startling theoretical conclusions regarding the compatibility of Riemann geometry
with Lorentz violation as well as proposals for new types of experiments.

The remainder of this section
addresses a number of key issues
associated with the structure of the SME.
A natural question at this stage
is to ask where the \cof\ could come from
and what other assumptions one might make about,
for example,
their spacetime dependence.
Section \ref{symbreak}
addresses this issue.
Often a part of the full SME
that is most relevant to a class of systems
becomes the focus of a given study.
A number of such limits of the SME
are commonly used
and these limits are reviewed in section \ref{limits}.
Many other theoretical issues associated with the structure of the SME
have also been considered.
Section \ref{theory}
highlights the scope of the questions that have been addressed.
A slightly older and somewhat more technical review
of SME work is also provided by  \cite{Bluh2006}.

\subsection{Symmetry breaking}
\label{symbreak}

The mechanisms by which \cof\
could arise in the SME can be divided into two classes:
explicit Lorentz-symmetry breaking
and spontaneous Lorentz-symmetry breaking.
Explicit Lorentz violation
is characterized by directly assuming nonzero \cof\
in the background,
while spontaneous symmetry breaking
assumes that the Lorentz violation arises
dynamically as a Lorentz-violating solution associated with a Lorentz-invariant action.
Here we provide some discussion of spontaneous breaking
and some comparisons with explicit breaking.

Spontaneous symmetry breaking typically occurs
when the low-energy solutions of a system do not respect a symmetry
that exists in the theory.
A classic example is provided by the case of a marble
initially placed on the unstable equilibrium 
at the central high point of a Mexican hat.
Any perturbation will cause the marble to drop to a lower energy configuration
in the brim of the hat breaking the initial axial symmetry of the situation.

In the Standard Model,
a potential analogous to the Mexican hat
is used to give a vacuum expectation value to the Higgs field,
spontaneously breaking $SU(2) \times U(1)$ gauge symmetry.
The process fills the spacetime vacuum with a scalar condensate
that affects fields coupling to the Higgs.
The spontaneous breaking of a global symmetry
results in massless Nambu-Goldstone modes,
while the breaking of a local gauge symmetry
results in massive gauge bosons,
the $W$ and $Z$ bosons in the case of the Standard Model.
Other particles coupling to the Higgs field
also receive a mass related to the vacuum expectation value.

If the field that acquires a vacuum value
is a vector or tensor object,
then the spacetime becomes filled with a vector or tensor condensate,
rather than a scalar as in the case of the Standard-Model Higgs.
Here the original action has Lorentz symmetry,
but the nature of the vacuum hides this symmetry at low energy.
The couplings of such a field to Standard-Model fields
generates the terms of the SME
in a manner analogous to the development of mass
for Standard-Model particles coupled to the Higgs field.
Hence in flat spacetime,
the implications of the couplings 
to the vacuum values associated with spontaneous Lorentz symmetry breaking
are the same as when explicit breaking is assumed.
However,
additional phenomenology may arise due to 
both the Nambu-Goldstone \cite{BluhKost2005}
and massive modes \cite{BluhFung2008}
associated with Lorentz-symmetry breaking.
For example,
the long-range interaction of the associated Nambu-Goldstone mode
can be identified with existing long-range interactions in nature,
or must be interpreted as a presently undiscovered interaction.
To date,
models generating the photon in Einstein-Maxwell theory \cite{BluhKost2005} 
and the graviton in General Relativity \cite{graviton}
as a consequence of Lorentz violation have been found.
Other types of new interactions have also been considered \cite{lvgap,AltsBail2010,NewInt}.

A class of vector theories 
having spontaneous symmetry breaking
known has bumblebee models
were used to initiate the investigation of spontaneous Lorentz violation \cite{KostSamu1989b}
and have recently been used to explore 
a variety of aspects of Lorentz-symmetry breaking 
\cite{BluhKost2005,BluhFung2008,lvgap,BailKost2006,RecentBumblebee}.
Additionally, there is considerable early literature
associated with couplings to vacuum-valued vector fields,
references to which can be found in section III A of  \cite{BluhFung2008}.
Similar models have also been used in a variety of contexts \cite{Beelike}.
Beyond vector theories,
spontaneous symmetry breaking in tensor theories
has also been considered \cite{graviton,AltsBail2010},
and some general results have been achieved \cite{tensor}.

While one is free to think of SME coefficients
in flat spacetime as existing either by virtue of explicit or spontaneous 
Lorentz violation,
incorporation of gravitation based on Riemann geometry
requires one to consider spontaneous breaking \cite{Kost2004}.
The result,
shown in the context of Riemann-Cartan geometry,
forms a no-go theorem for explicit Lorentz breaking
in gravity.
The basic idea is that the structure of Riemann-Cartan geometry
implies that the Bianchi identities of the curvature
along with the field equations
provide a constraint on the covariant conservation laws.
This constraint is generally not satisfied for explicit breaking,
signaling that such a theory is not self consistent.
Conversely the constraint is automatically satisfied for spontaneous Lorentz violation
due to the fact that the Lorentz-violating fields arise dynamically from within the theory.
Reference \cite{Kost2004} provides
additional discussion and examples.
The constraint that arises is developed explicitly in section \ref{lo_riemann}
for the case of the leading-order Riemann limit of the pure gravity sector.

The no-go theorem for explicit Lorentz violation in gravity
is a very significant result for several reasons.
It provides one of the few opportunities to make a general statement
about the nature and origin of possible Lorentz violation:
if Lorentz-symmetry is violated, it must either occur spontaneously
or a new geometrical framework for gravity must be found.
As noted above,
spontaneous breaking is distinguished from explicit breaking
by the existence of modes related to fluctuations
about the vacuum values in addition to the vacuum values themselves.
The no-go theorem implies that consistent work with Lorentz violation in gravity
requires consideration of the fluctuations in addition to the vacuum values
associated with Lorentz breaking.
This requirement makes gravitational phenomenology considerably more challenging.
Model-independent methods of addressing the fluctuations 
have been achieved in phenomenological studies of the pure-gravity sector \cite{BailKost2006}
as well as the gravitationally coupled matter sector \cite{lvgap}.
The incompatibility of Riemann geometry
and explicit Lorentz violation
has also spurred consideration of Finsler geometry \cite{Kost2011,finsler}.
Consideration of the geometric structure
associated with SME coefficients has led to the discovery
of new Finsler spaces known as SME spaces \cite{Kost2011}.
Other resolutions to the question of geometrical consistency may be possible
if a suitable nongeometrical theory of gravity with Lorentz violation \cite{horava_etal} were found.
The no-go theorem also has implications
for going beyond conventional theory \cite{rb2014}.

\subsection{Popular limits of the SME}
\label{limits}

The full SME
includes an infinite number of operators
of ever-increasing mass dimension.
While this might seem daunting,
several approaches have been used to keep the process tractable.

\subsubsection{Minimal SME}
\label{minimal}

The most commonly used limit of
the SME is the {\it minimal SME} \cite{CollKost1998},
which has been studied extensively in the literature
both theoretically and experimentally.
This limit could be regarded
as containing the leading Lorentz-violating terms
of the series-approximation vision
highlighted in section \ref{sec:sme}.
In this limit,
basically all of the usual properties of the Standard Model
are maintained except particle-Lorentz symmetry
and CPT symmetry.
To be explicit,
the following properties are maintained:
the usual $SU(3) \times SU(2) \times U(1)$
gauge structure,
power-counting renormalizability,
energy and momentum conservation,
$SU(2) \times U(1)$ symmetry breaking,
quantization,
microcausality,
spin-statistics,
and observer Lorentz covariance.
Here maintaining energy and momentum conservation
implies that attention is restricted
to constant \cof.
That is,
\beq
\label{ConsCoef}
\prt_\al t_{\mn \la \ldots} = 0,
\eeq
for any coefficient $t_{\mn \la \ldots}$.
This assumption is reasonable since it could be regarded
as the leading term in a Taylor expansion of 
a coefficient with spacetime dependence.
Additionally, many SME studies assume the \cof\
are perturbatively small,
based on the fact that Lorentz violation
has not yet been seen in nature.
Additional properties of specific coefficients
are summarized in  \cite{data}.
The assumptions of the minimal SME
focus attention on one deviation from known physics,
in this case
Lorentz violation and its natural consequences.
This reflects a principle frequently applied
in SME studies,
which is sometimes referred to as `Kosteleck\'y's Cutlass'.
The principle states that no more than one deviation from known physics
should be considered a time.
The idea being that if $\ep$ is the probability
that a given deviation from known physics provides a correct description of nature,
then the probability of two such unobserved deviations being found together
is order $\ep^2$.
Kosteleck\'y's Cutlass can then be used to cut many such suggestions
from consideration.

An associated limit
is the gravitationally coupled minimal SME \cite{Kost2004}.
Here minimal couplings to gravity
are considered in the same terms that appear
in the flat-spacetime SME.
In this context,
asymptotically Minkowski spacetimes are often considered
in which
 \rf{ConsCoef}
is assumed to hold asymptotically,
implying energy and momentum conservation
asymptotically.
However,
geometrical consistency 
typically prevents the application
of this condition beyond the asymptotic limit.

\subsubsection{QED extension}

In much the way quantum electrodynamics (QED)
can be extracted from the Standard Model,
the QED extension can be extracted from 
the Standard-Model Extension \cite{CollKost1998}.
The minimal QED extension
is a popular limit
studied extensively in the literature.
The associated fermion lagrangian can be written
\beq
\label{Lps}
\cl_\ps =  \frac{1}{2} i \ol{\ps} \Ga_\nu \lrvec{D^\nu} \ps 
- \ol{\ps} M \ps,
\eeq
where
\beq
\Ga_\nu \equiv \ga_\nu + c_\mn \ga^\mu + d_\mn \ga_5 \ga^\mu 
   + e_\nu + i f_\nu \ga_5
   + \half g_{\la \mu \nu} \si^{\la \mu},
\eeq
and
\beq
M \equiv m + a_\mu \ga^\mu + b_\mu \ga_5 \ga^\mu 
   + \half H_\mn \si^\mn.
\eeq
Here \a, \b, \c, \d, \e, \f, \g, and \H\
are fermion-sector coefficients for Lorentz violation,
$\ps$ is the fermion field,
and the $\ga^\mu$ are the usual Dirac matrices.
The covariant derivative in this context
is now $D_\mu = \prt_\mu + i q A_\mu$,
where $A_\mu$ is the photon field.
Setting the coefficients to zero
results in the Lorentz-invariant limit
reproducing the conventional Dirac lagrangian.
It is common to treat mesons and baryons
as having independent coefficients for Lorentz violation
in this context.
Though difficult in practice,
these coefficients could be expressed in terms
of the quark and gluon content of the particle.

The photon lagrangian takes the form
\bea
\nonumber
{\cl}_{A} 
&=& 
-\frac 14 F_{\mu\nu}F^{\mu\nu}
-\frac 14 (k_F)_{\ka\la\mu\nu} F^{\ka\la} F^{\mu\nu}\\
& & 
+ \half (k_{AF})^\ka \ep_{\ka\la\mu\nu} A^\la F^{\mu\nu}
- (k_A)_\ka A^\ka,
\label{La}
\eea
where the coefficients for Lorentz violation here
are $(k_F)_{\ka\la\mu\nu}$,
$(k_{AF})^\ka$,
and $(k_A)_\ka$,
and $F_\mn \equiv \prt_\mu A_\nu - \prt_\nu A_\mu$
is the field strength.
A helpful analogy to the effect of the \cof\
is provided by electrodynamics in anisotropic 
and gyrotropic media \cite{CollKost1998}.
Note also that the Chern Simons term \cite{CarrFiel1990},
which continues to receive considerable attention \cite{chern}
is contained here in the $k_{AF}$ term.

As a means of providing a sense of the number of \cof\
involved a typical SME analysis,
we consider the number of 
independently observable coefficients associated
with ordinary matter
(protons, neutrons, and electrons),
in the nonrelativistic flat-spacetime limit
of the QED extension. 
Here there are 132 independently observable coefficients \cite{data}.
Based purely on a counting of coefficients,
this limit is comparable 
in complexity to the number of free parameters
in the minimal supersymmetric model,
another popular deviation from known physics.
Note however,
that the spirit of the \cof\ in the SME is to
provide a broad catalogue of all possible deviations from Lorentz symmetry,
which is rather different from the free parameters of a 
specific model such as supersymmetry.

To gain further intuition
about the coefficients for Lorentz violation,
consider two examples.
First,
in the newtonian limit,
the \c\ coefficient generates the direction-dependent effective inertial mass
\cite{Bert}
considered in section \ref{rotation}
with
\beq
m_{jk} = m (\de_{jk} + c_{jk} + c_{kj}).
\label{eq:mjk}
\eeq
The implications of \c\ at any scale are similar,
altering the relation between speed and energy
as well as
velocity and momentum,
and doing so in a direction-dependent way.

As another simple example,
consider the coefficient \b,
having an axial vector coupling to the fermion field above.
The leading effect of this coefficient at the nonrelativistic level \cite{KostLane}
is the contribution 
\beq
H_{\rm Non \ Rel} \supset \bi{b} \cdot \bi{\si},
\label{eq:bdotsi}
\eeq
where $\bi b$ is the spatial content of \b.
This term leads to a precession of spins \cite{BluhKost2000},
an effect that has been studied extensively,
which reflects the violation of angular-momentum conservation
that accompanies Lorentz violation.

The gravitationally coupled minimal QED extension
has also been considered \cite{Kost2004,lvgap}.
Here minimal couplings to gravity
are considered in the same terms appearing in Eqs.\ \rf{Lps} and \rf{La}.
The vierbein formalism is used for the incorporation of fermions
and the derivatives become covariant derivatives for spacetime
as well as $U(1)$.
As an example,
the term associated with the \c\ coefficient
takes the form
\beq
\cL_\ps 
\supset
\half i e \ivb \mu a \ol \ps 
c_{\al\be} \uvb \be a \ivb \al b \ga^b
 \lrDmu \ps 
\label{qedxps},
\eeq
where 
$\vb \mu a$
is the vierbein
and
$e$ is the vierbein determinant.
As is standard in SME studies with gravity,
Greek indices refer to general spacetime coordinates,
while Latin indices refer to local Minkowski coordinates.

\subsubsection{Leading-order Riemann limit}
\label{lo_riemann}

A key condition of the minimal SME
is power-counting renormalizability,
a property already absent from conventional General Relativity.
The analogous limit
in the gravity sector
is the restriction to operators of no higher mass dimension
than those of the conventional Einstein-Hilbert action.
Such terms are referred to as leading-order terms.
A further restriction
that is natural in many circumstances
is the Riemann limit
in which torsion vanishes.
This limit is relevant,
for example,
in solar-system phenomenological studies
where torsion effects are typically suppressed
making coupling of \cof\ to torsion
of considerably lesser interest than
couplings of Lorentz violation
to curvature.
It is also common to specialize further
to the limit of linearized gravity \cite{BailKost2006,bailey2010}.

Considerable work
on Lorentz violation in gravity has also been done
in the context of the parametrized post-newtonian (PPN) formalism \cite{Will}.
The philosophy of the PPN
is somewhat analogous to the philosophy of the SME;
however,
their goals and methods are rather different.
Reference \cite{BailKost2006}
provides a detailed comparison between
the PPN
and the leading-order Riemann limit
of the SME.
Here we summarize several key points.
As a comparison of motivation,
that of the PPN
is to provide a test framework
parameterizing deviations from General Relativity,
some of which are Lorentz violating,
while the SME parametrizes deviations from exact Lorentz symmetry,
some of which result in deviations from General Relativity.
In terms of methods,
the PPN
provides an expansion about the General Relativity metric,
while the SME provides an expansion about
the action of General Relativity and the Standard Model.
In terms of Lorentz violation,
the PPN assumes that physics is isotropic
in a particular frame,
while the SME makes no such assumption.
Perhaps most interestingly,
reference \cite{BailKost2006}
finds only a one degree of freedom overlap between
the PPN
and the leading-order Riemann limit
of the SME,
implying that the SME provides many new opportunities
for existing gravitational experiments.

Since the pure-gravity sector is an area where the no-go theorem of section \ref{symbreak}
plays an especially prominent role,
we conclude this section by illustrating its implications
in the leading-order Riemann limit
of the SME.
The pure-gravity action in this limit takes the form
\beq
S_{\rm gravity}
=
 \fr 1 {2\ka}\int d^4 x 
[ e(1-u)R -2e\La
+ e s^\mn R_{\mu\nu}
+ e t^{\ka\la\mn} R_{\ka\la\mu\nu}
].
\label{Ract}
\eeq
Here $1/2\ka \equiv 1/16\pi G_N$, where $G_N$ is Newton's constant,
$R$ is the curvature scalar,
$R_{\mu\nu}$ is the traceless Ricci tensor,
$R_{\ka\la\mu\nu}$ is the Weyl tensor,
$\La$ is the cosmological constant,
$s^\mn$ and $t^{\ka\la\mn}$ are coefficient fields for Lorentz violation,
and $u$ is a coefficient field though not Lorentz violating.
Variation of the action
with respect to the metric $g_\mn$
yields a modified Einstein equation of the form
\beq
G^\mn - (T^{Rst})^\mn = \ka T_g^\mn.
\label{ein}
\eeq
Here the material on the left comes from variation of the partial action $S_{\rm gravity}$
from equation \rf{Ract},
where $G^\mn$ is the usual Einstein tensor and $(T^{Rst})^\mn$
contains the additional material associated with the coefficient fields in equation \rf{Ract}.
In a theory having spontaneous breaking, 
there will also be a partial action associated with the dynamics
of the coefficient fields
that may contribute energy momentum to the term on the right-hand side of \rf{ein}.
In the present context of pure gravity only,
this will be the only contribution to the right-hand side.
Acting on \rf{ein} with $D_\mu$ 
and using the trace Bianchi identity $D_\mu G^{\mu\nu} = 0$
yields
\beq
0 = D_\mu (\ka T_g^\mn + (T^{Rst})^\mn),
\eeq
imposing a constraint on the coefficient fields contained within $(T^{Rst})^\mn$.
The trace Bianchi identity stems purely from geometry,
resulting in a geometric constraint on the energy momentum tensor
that arises through variation with respect to $g_\mn$.
This constraint is not in general satisfied for explicit Lorentz-symmetry breaking,
but is automatically satisfied for spontaneous-symmetry breaking.
This can be verified explicitly in specific models.
Additional discussion and examples
appear in reference \cite{Kost2004}.

\subsubsection{Nonrenormalizable operators}
\label{sec:nro}

A natural way to go beyond the minimal SME
is to relax the condition
of power-counting renormalizability,
and hence consider Lorentz-violating operators of arbitrary mass dimension.
This allows consideration
of the full Lorentz-violating series approximation
of the underlying theory.
These higher-dimension operators
might be expected to be particularly relevant
in very high energy processes.
Models containing Lorentz violation also exist
in which higher-dimension operators
provided the leading \cite{honc}
or dominant Lorentz-violating effects \cite{Berg2002,hoss}.

Nonrenormalizable operators
of arbitrary mass dimension have been considered explicitly and comprehensively
in the pure photon sector \cite{KostMewe2009}
and in the fermion sector \cite{KostMewe2012,KostMewe2013}.
More specialized work on higher-dimension Lorentz violation
has also been done \cite{MyerPosp2003,propgammamatter,holv}.
In the same spirit as the minimal SME,
references \cite{KostMewe2009,KostMewe2012,KostMewe2013} consider nonrenormalizable Lorentz-violating terms
that maintain most of the other usual properties of the Standard Model.
To do so,
they focus on operators that are quadratic in the conventional fields
and that maintain conservation of energy,
momentum,
and electric charge.
The subset of operators having these properties
can typically be written in a form
similar to terms in the minimal SME
in which the objects appearing in a manner
analogous to the minimal-SME terms
become differential operators.

Here we consider
the infinite set of CPT-odd coefficients
of mass dimension $d$
in the photon sector
as an example of the above structure.
The contribution of these terms to the lagrangian
takes the form
\beq
\cl \supset \frac 1 2 \ep^{\ka\la\mu\nu}A_\la (\kaf)_\ka F_{\mu\nu},
\label{nmphoton}
\eeq
where
\beq
(\kaf)_\ka =\hspace{-3pt}\sum_{d=\mbox{\scriptsize odd}} 
{(\kafd{d})_\ka}^{\al_1\ldots\al_{(d-3)}} 
\prt_{\al_1}\ldots\prt_{\al_{(d-3)}}.
\eeq
The sum here ranges over values
$d\geq 3$.
Note that  \rf{nmphoton}
differs from  \rf{La}
only by the hat 
above the $(\kaf)_\ka$
indicating that it is now a differential operator
rather than a constant coefficient as in the minimal case.
Note also that the minimal case
is recovered in the limit $d=3$.

The infinite number of coefficients studied
in the nonrenormalizable limit
are further tamed by a classification scheme 
based on the rotation properties of the coefficients.
The method results in an expansion 
in terms of spherical coefficients for Lorentz violation
and spin-weighted spherical harmonics.
Such a decomposition is well suited to many experimental scenarios,
which take advantage of the rotation properties
of the coefficients.

\subsection{Theoretical work}
\label{theory}

A large amount of work has been done on various theoretical properties
of the SME.
Though space prohibits a full consideration of many of these fascinating topics,
this section briefly highlights some of the areas that have been explored
and provides references for the interested reader.

\subsubsection{Observability of coefficients}
In a few cases,
coefficients for Lorentz violation
appearing in the general SME expansion
are unobservable
and can be removed from the theory
by field or coordinate redefinitions.
As one simple example of the required style of thinking,
consider the \a\ term appearing in the fermion sector \rf{Lps}.
The field redefinition,
$\ps(x) = \exp[i a_\mu x^\mu]\ch(x)$,
will remove this term from the theory in the single-fermion Minkowski-spacetime limit \cite{CollKost1998}.
Note that this is a field redefinition and not a gauge transformation,
as the field $A_\mu$ is uninvolved.
If one instead chooses to work with the original field $\ps$,
\a\ terms will remain in relevant calculations,
but will not lead to observable effects in the single-fermion Minkowski-spacetime limit.
The \a\ coefficient can be observed outside of this limit.
References \cite{CollKost1998,Kost2004,KostMewe2009,KostMewe2013,lvgap,KostMewe2002,KostMewe2008,observability}
consider various other instances in which coefficients can be removed from the theory.

\subsubsection{Stability and causality}
The questions of stability and causality in 
the SME were first considered in detail in Ref.\ \cite{KostLehn2001}.
This work considers the question
in the context of the single-fermion limit of the free-matter sector
of the SME.
It is found that difficulties with stability or causality generally
arise in theories having explicit Lorentz violation;
however,
if the coefficients are Planck suppressed,
as expected in the typical motivation for the study of Lorentz violation,
the difficulties occur at high energies or high boosts only.
In this regime,
the validity of low-energy effective field theory might be questioned anyway.
This work also shows explicitly that spontaneous Lorentz violation
in suitable scenarios can avoid these problems.
In these scenarios,
nonrenormalizable terms play a key role as energies approach the Planck scale,
perhaps providing some bottom-up motivation for viewing the SME
as a series approximation for Planck-scale physics,
and providing support for the stability and causality of the SME
as a theory emerging at low energies from spontaneous breaking 
in a realistic string theory.

Work on stability and causality has also been done
in the context of other areas of Lorentz violating quantum field theory \cite{stability}.
For example,
a scalar quantum field theory with Lorentz violation
has been shown to have reasonable stability and causality properties \cite{altschul2005}.

\subsubsection{Renormalization}
While the minimal SME contains operators
that are power-counting renormalizable,
the details of renormalization have been the subject
of considerable work.
Here we highlight some key results
and demonstrate the breath of work that has been done.
Early work demonstrated the 1-loop renormalizability of the QED extension \cite{KostLane2002}.
Renormalizability at 1 loop has since been shown for pure Yang-Mills theory
with Lorentz violation \cite{cm2007renorm},
a result subsequently extended to include fermions, along with some additional generalization of results,
to show the explicit 1-loop renormalizability of the gluon sector of QCD with Lorentz violation
\cite{cm2008renorm}.
The 1-loop renormalizability of the electroweak sector of the SME has also been investigated
\cite{cm2009renorm}.
Renormalization in Lorentz-violating QED
beyond 1 loop \cite{delcima2012},
and in a curved background \cite{berredo2006}
have been considered,
as have topics including scalar and Yukawa field theories \cite{ferrero2001}
and nonpolynomial interactions \cite{altschul2005a}.

In addition to establishing basic features of the field theory,
the associated question of the running of the coefficients for Lorentz violation
is relevant to understanding the role of Lorentz violation across energy scales
\cite{KostLane2002}.

A variety of other quantum field-theoretic properties 
have also been explored in the context of Lorentz violation.
For example,
the following areas have been explored:
radiatively induced Lorentz violation \cite{radiative},
Yang-Mills instantons \cite{colladaymcdonald2004},
properties of the modified Dirac equation \cite{lehnert2004},
the connection between noncommutative field theory and the SME \cite{carroll2001},
the K\"all\'en-Lehmann representation for Lorentz-violating field theory \cite{potting2012},
Gupta-Bleuler photon quantization \cite{colladay2014},
massive photon theory \cite{cambiaso2012},
Yukawa-type quantum field theory 
with Lorentz violation in the bosonic sector \cite{altschul2006},
theories with nonpolynomial interactions and spontaneous Lorentz violation 
\cite{altschulkost},
and the implications of spontaneously breaking gauge symmetry with Lorentz violation
\cite{altschul2012}.

\subsubsection{Modified reactions}
Beyond technical questions associated with the SME
as a field theory,
considerable work has been done associated
with calculating the nature of modifications
to various high-energy processes.
Here we list some examples of work in this area
that the interested reader may wish to explore.

Early work on the SME established some general features of cross sections and decay rates \cite{CollKost2001}.
In the usual analysis of some processes,
Lorentz invariance is used in developing the necessary tools.
Some of the alternative procedures needed were developed in this work.
Electron-positron pair annihilation into two
photons was considered as a specific example.
Some other examples of modified processes considered in the literature
include:
Compton scattering \cite{Alts2004},
synchrotron radiation \cite{Alts2005},
Lorentz-violating effects on the thresholds of various processes \cite{Lehn2003},
the possibility of photon splitting in Lorentz-violating field theory \cite{photonsplit},
modifications to $\al$-decay associated with the composite coefficients for Lorentz violation
of the $\al$ particle \cite{Alts2009},
and
the effects of CPT violation on baryogenesis \cite{Bert1997}.
In some cases,
processes forbidden in conventional physics become allowed.
Vacuum \v Cerenkov is one example that has been considered fairly extensively in the literature \cite{cerenkov}.
The effect becomes allowed as charged particles may exceed the vacuum speed of light
in the presence of some coefficients for Lorentz violation.
There is a strong analogy here with conventional \v Cerenkov radiation in materials.

\subsubsection{Supersymmetry}
Connections with other symmetries have been studied
including various investigations of connections with supersymmetry.
In reference \cite{Berg2002},
it was established that Lorentz-violating supersymmetric field theories exist.
This was illustrate with simple examples related to the Wess-Zumino model.
There is also a philosophical connection between supersymmetry and Lorentz violation highlighted in that work.
Supersymmetry is a hypothesized spacetime symmetry.
If it exists in nature, 
experiment suggests that it must be broken,
and spontaneous breaking is an attractive mechanism.
This situation seems somewhat in parallel with the present discussion of the possible
breaking of the spacetime symmetries of Lorentz and CPT symmetry.
The interested reader can find more information about supersymmetry and Lorentz violation
in references \cite{hoss,supersymmetry}.
As noted in section \ref{sec:nro},
another interesting feature of some supersymmetric models 
is the generation higher mass-dimension Lorentz-violating terms
as the dominant Lorentz-violating contribution \cite{hoss}.

\subsubsection{Classical and nonrelativistic limits}
While the SME is a relativistic quantum field theory,
various limits such as the classical or nonrelativistic limits
yield additional theoretical understanding
and are useful for many phenomenological investigations.
Various aspects of these limits have also received considerable attention \cite{Bert,KostLane,classicalnr}.
Examples of these uses have been considered above
in Eqs.\ \rf{eq:mjk} and \rf{eq:bdotsi}.

\section{Experiments and observations}
\label{sec:tests}

This section addresses basic aspects of searches
for Lorentz violation in the SME,
summarizes the types of experiments that have been done,
and offers some comments on current constraints.

\subsection{Standard frame}

Under an observer Lorentz transformation,
the form of the \cof\ changes.
For example,
an observer boost by $\be$ in the $x$
direction of some observer frame
results in 
\bea
\nonumber
b_{0^\prime} &=& \ga (b_0 - \be b_x)\\
b_{x^\prime} &=& \ga (b_x - \be b_0).
\eea
Hence constraints on $b_0$ and $b_x$
obtained in the original observer frame
will appear as mixed and scaled constraints
on $b_{0^\prime}$ and $b_{x^\prime}$
when reported in the new frame.
Additionally,
the \cof\ will become time dependent
in a frame that is constantly rotating or changing its boost.

\begin{figure}
\begin{center}
\includegraphics[scale=0.25]{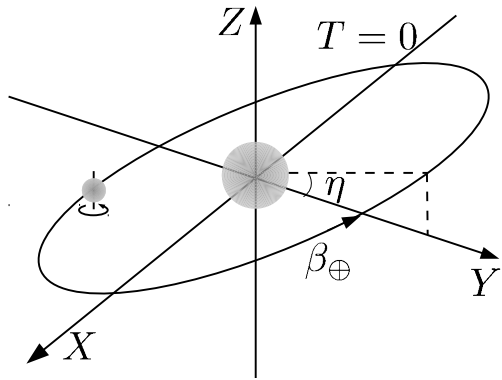}
\caption{Sun-centered frame \cite{data}.}
\label{fig1}
\end{center}
\end{figure}

To address these issues in a manner most relevant for experiments
on or near Earth,
the Sun-centered celestial equatorial frame 
has been chosen as the standard frame
for reporting measurements of \cof\ \cite{Bluh2002}.
Figure \ref{fig1}
illustrates this frame.
The vernal equinox in the year 2000
is chosen as the origin of the time coordinate, $T$.
The $Z$ axis is parallel to the rotation axis of the Earth
at $T=0$,
the $X$ axis points from the Sun toward the vernal equinox,
and the $Y$ axis completes the right-handed coordinate system.
Note that the choice of capital letters
for the coordinates in this system
is standard in the literature.
Here $\be_\oplus$ is the boost velocity of the Earth in the Sun-centered frame
having a value of approximately $10^{-4}$
and $\et \approx 23.5^o$.
Additional information about this frame
including transformations to other standard frames
can be found in Section III A and Appendix C of  
\cite{KostMewe2002},
while implications of the frame choice in a gravitational context
are considered in \cite{lvgap,BailKost2006}.
This choice of standard frame
is useful since it is approximately inertial over the time-scales
of most relevant experiments,
and it is convenient for the majority of tests
performed on Earth and in the solar system.
It is also worth noting that this frame is chosen
simply to provide a standardized choice for reporting constraints.
This frame is not a preferred frame
as the term is often used in the literature,
nor are any other special assumptions made about it.

\subsection{Experimental and observational signatures}

In searching for,
or attempting to constrain Lorentz violation,
experiments and observations typically take advantage
of one or more of the following basic ideas:
classic boost or orientation dependence,
particle species dependence,
CPT testing,
or modified processes.

The experimental or observational hallmark of Lorentz violation
is a result that changes when the experiment is 
repeated in a new inertial frame having a new orientation
or new velocity.
In the original Michelson-Morley experiment,
as well as modern-day improvements,
this would correspond to a fringe shift
as the interferometer is rotated in the lab.
Another modern test of Lorentz symmetry
that provides a popular example
is to search for a variation in the tick rate of a clock
as the system is rotated or boosted \cite{KostLane1999}.
Many modern experiments use rotation in the lab,
but many more take advantage of the rotation of the experiment provided by the Earth
over a sidereal day.
Boost invariance can also be tested
as experiments in the lab change velocity do the Earth's sidereal rotation
and annual revolution.
Rotating satellites 
and those with higher boost factors can provide even more advantages \cite{lvgap,Bluh2002}.

Other types of signatures
take advantage of the possibility of particle-species dependent
coefficients for Lorentz violation.
As one example,
effective Weak Equivalence Principle violation results
when terms involving particle-species dependent fermion-sector coefficients
are coupled to gravity.
This results in novel signals in Weak Equivalence Principle tests
that come with the characteristic boost and orientation dependence of Lorentz violation \cite{lvgap}.

Another key signal arises in comparing particles and antiparticles.
As discussed above,
coefficients with odd numbers of indices are CPT odd
while those with even numbers of indices are CPT even.
This implies different predictions for antiparticles
over particles as CPT odd terms change sign.
Under special circumstances tests with antimatter
can yield novel sensitivities to \cof\ \cite{lvgap}.
As an example,
differences in the spectrum of hydrogen and antihydrogen
would be a signal of CPT violation,
which would also come with the
characteristic boost and orientation dependence of Lorentz violation \cite{BluhKost1999}.

Finally,
the additional terms included in the SME
can lead to modifications to various physical processes,
which would also come with boost and orientation dependence.
A simple example is provided by the rotation-invariance violation
that enters Newton's second law
from SME coefficient \c\ discussed in section \ref{rotation}.
The orientation dependent effective mass here
leads to accelerations in directions in which there is no force.
As a more exotic example,
Lorentz violation in the photon sector
can lead to vacuum birefringence.
Searching for such possibilities
often offers the possibility of impressive observational
sensitivities.

\subsection{General considerations}
\label{generalconsiderations}

Before discussing sensitivities achieved via the above methods,
it is worth considering a few general aspects
of the search for Lorentz violation.
First,
it should be noted that there is really no such thing
as a single ``best test of Lorentz symmetry''
or ``best test of CPT symmetry''.
Imagine the analogous statement
about parity symmetry
prior to the discovery of parity violation.
Claiming the best test of parity
in electromagnetic interactions
is irrelevant in claiming that other interactions are parity invariant
and such a test
does nothing to help detect the parity violation
that does exist in nature.
Similarly,
it is not really possible to make
particularly compelling statements about which
\cof\
are most likely to be nonzero,
or which experiments have the best probability
of finding Lorentz violation.
A constraint of $10^{-40}$ 
with photons \cite{data} does not exclude a signal at $10^{-10}$
in gravity.

It is also tempting ask
at what level one might expect to find
Lorentz violation originating from the Planck scale.
Any attempt at a definitive answer to this question
would likely be highly model dependent
and at odds with the philosophy of a broad search.
However,
dimensional arguments provide
some sense of scale.
For example,
in the case of a dimensionless coefficient,
one might imagine Planck-suppressed effects
generating nonzero values
on the order of the ratio of an energy scale associated
with the relevant physics to the Planck mass.
Hence one could consider,
for example,
the mass of the electron over the Planck mass
and arrive at $10^{-23}$
or perhaps at the larger extreme,
the electroweak scale over the Planck scale
and arrive at $10^{-17}$.

\subsection{Current limits}
\label{currentlimits}

To date,
no compelling evidence for Lorentz violation has been found.
Thus this section focuses
on current limits
and large areas of open space for exploration.
Current limits on \cof\ are tabulated 
and updated annually in the publication
{\it Data Tables for CPT and Lorentz Violation} \cite{data}.
At the time of writing this review,
there are over 1000 published limits
tabulated in that work
(including multiple improvements in sensitivity
to the same coefficient).
Though it is not appropriate,
nor is there space,
to repeat all of that material here,
this subsection will summarize
the breath of measurements that have been made,
the impressive sensitivity has has been reached
in some cases,
and the large unconstrained regions of coefficient space.

Sensitivities have been achieved with a large number
of particles using a wide variety of physical systems.
These include sensitivity to coefficients
in the minimal SME
associated with
electrons \cite{CarrFiel1990,KostLane1999,npesensitivity,pesensitivity,nesensitivity,egammasensitivity,Alts2007,Hohe2013,esensitivity},
protons \cite{KostLane1999,npesensitivity,pesensitivity,npsensitivity,psensitivity},
neutrons \cite{KostLane1999,npesensitivity,nesensitivity,npsensitivity,comag,nsensitivity},
photons \cite{KostMewe2009,KostMewe2002,KostMewe2008,egammasensitivity,Caro2006,KostMewe2007,gamma,gammasensitivity},
muons \cite{Alts2007,musensitivity},
taus \cite{Alts2007},
neutrinos \cite{KostMewe2012,BargMarf2007,CoheGlas2011,nusensitivity},
quarks \cite{Alts2007,qsensitivity},
the Higgs \cite{phsensitivity},
the $W$-boson \cite{Alts2007},
and
the gluon \cite{Caro2006}.
Coefficients associated 
with protons, neutrons, and electrons
have also been sought using gravitational couplings \cite{jt2012,lvgap,Hohe2013,mgsensitivity}
and tests have been used to place constraints
on coefficients associated with the gravitational field
in the leading-order Riemann limit \cite{BailKost2006,gravitysensitivity}.
Nonrenormalizable coefficients
associated with photons \cite{KostMewe2009,KostMewe2007,gamma,nrgammasensitivity},
neutrinos \cite{KostMewe2012},
and other fermions \cite{KostMewe2013}
have also been investigated experimentally and observationally.
Fermion results based on a number of works \cite{nrfermi}
are tabulated in \cite{KostMewe2013}.
In addition to the sensitivities noted above,
considerable phenomenological work suggesting tests
has also been done.
Examples of these proposals are contained within many of the references
noted above that contain some constraints
as well as in a number of theoretical references.
Here we point the reader
to several key phenomenological works
in various areas
as well as to some phenomenological analysis
not cited elsewhere in this work:
photons \cite{KostMewe2009,propgammamatter,KostMewe2002,KostMewe2007,propgamma};
space-based tests \cite{Bluh2002,propspace};
neutrinos \cite{KostMewe2012,KostMewe2004,DiazKost,DiazKost2013,propnu};
ordinary-matter tests such as clocks, particle motion, etc.\ \cite{KostMewe2013,lvgap,KostLane1999,propordi};
antimatter \cite{lvgap,BluhKost1999,propanti};
mesons \cite{propmeso};
and
gravity \cite{lvgap,BailKost2006,propgrav}.

If the arguments
of section \ref{generalconsiderations}
provide a reasonable guide,
experiments are in some cases
approaching the astounding sensitivity
required to probe effects at the level of 2 Planck suppressions
and some observational sensitivities have exceeded this level.
For example,
sensitivity to the \b\ coefficient for the neutron has reached the level of $10^{-32}$ GeV
in comagnetometer experiments \cite{npsensitivity,comag}
and sensitivities to $\bi{k_{AF}}$ in the photon sector
have reached the level of $10^{-43}$ GeV via CMB polarization analysis \cite{KostMewe2009,KostMewe2008,KostMewe2007}.
At the other end of the spectrum,
large regions of coefficient space remain open,
with no explicit observational or experimental constraints.
For example,
only one constraint exists associated with the tau lepton,
and even in the context of minimal coefficients associated with ordinary matter,
one of the most accessible places for experiments,
dozens of coefficients remain unconstrained.
Thus a large number of ways Lorentz symmetry could be violated remain unexplored.
In some cases Lorentz violation could still be comparatively large
and yet have evaded detection to date,
a possibility known as countershaded Lorentz violation \cite{lvgap}.
Matter-sector coefficients that are observable only when coupled to the gravitational field \cite{lvgap}
and neutrino-sector coefficients observable only in processes involving
neutrino phase-space properties \cite{DiazKost2013} are explicit examples.
In such cases,
Lorentz violation can be large enough that the Lorentz hierarchy problem \cite{KostPott1995}
can be obviated.

As a final note,
although no compelling evidence for Lorentz violation has been found,
it is worth noting that one cannot quite call
all of the experimental and observational work that has been done
``limits''
since a few tests find nonzero values at the level of a few sigma.

\subsection{Neutrinos}

Neutrinos deserve special comment in the context of Lorentz violation
for several reasons.
First,
neutrinos are perhaps the only area 
in which physics beyond the Standard Model 
is seen quite definitively
in the form of neutrino oscillations.
While massive neutrinos are a natural way to explain the observed oscillations,
the frequently-heard statement that oscillations imply massive neutrinos
is not correct.
In fact it is possible to generate oscillations
in SME-based models with Lorentz violation
and massless neutrinos \cite{KostMewe2004,BargMarf2007,massless}.
Second, several pieces of experimental evidence
do not fit within the 3 mass model
such as the MiniBooNE anomaly \cite{miniboone},
the LSND anomaly \cite{LSND},
and CPT asymmetries of the MINOS type \cite{MINOS}.
These effects can be accommodated in some SME-based models \cite{KostMewe2012}.
Other SME-based models for neutrinos have also been developed \cite{DiazKost,numodels}
having a variety of features.
For a general discussion of classes of models,
see Section V of  \cite{KostMewe2012}.
In addition to efforts to build models with Lorentz violation
as alternatives to the standard 3 mass model,
one can also consider Lorentz violation as a perturbing effect
on the standard 3 mass model.

Finally,
the possibility of neutrinos as faster-than-light particles
has been around for some time \cite{ChodHaus1984}
and is an effect that can arise due to nonzero SME \cof.
The idea received considerable attention
due to the recent OPERA result indicating such an observation \cite{OPERA1},
now believed to be a systematic \cite{OPERA2},
of neutrinos traveling faster than the speed of light.
At first glance,
this was perhaps the most exciting piece of experimental information
to arise in the field of Lorentz violation.
Such a result would have provided clear evidence of Lorentz violation
interpretable as nonzero SME \cof.
On closer inspection,
the result had several surprising features
that perhaps further increased the degree of skepticism
that such a paradigm-changing result naturally would receive.
First,
the effect was quite large,
with neutrinos appearing to exceed the speed of light by $10^{-5}$ \cite{OPERA1},
an effect that could occur with,
for example,
dimensionless SME coefficients of the same order \cite{KostMewe2012}.
While it has been shown that Lorentz violation of this size could certainly
evade detection in special circumstances \cite{lvgap},
the effect is large compared with Planck suppression
as well as many existing limits.
The size of the signal also made it more challenging
to incorporate the result alongside
other time of flight measurements
and other data on Lorentz violation in neutrinos \cite{data}.
Perhaps more interestingly,
one would typically expect vacuum \v Cerenkov radiation
to occur for faster than light particles \cite{cerenkov}
that would have an effect on the energy spectrum of the neutrinos;
however,
such an effect was not seen \cite{CoheGlas2011}.
If such a signal were observed in neutrinos,
the power of the SME as a complete effective field theory
would permit the testing of such an effect
in other processes
involving the same \cof.
The charged pion decay rate
would have been a useful candidate \cite{Alts2011}.

\section{Conclusions}
\label{conclusion}

In this work,
we have briefly reviewed the history of Lorentz symmetry
and its role in modern physics.
Some pedagogical examples have been provided
to illustrate the meaning of symmetry in physics
as well as how a violation of spacetime symmetry would look,
and the connection between Lorentz symmetry
and CPT symmetry has been presented.
The topic of Lorentz violation has been studied extensively
through theoretical, phenomenological, experimental, and observational work
in the context of the effective field-theory framework
of the SME.
Following discussion of the motivation for such a framework,
we consider the construction of the SME,
its popular limits,
and theoretical investigations of Lorentz violation.
The final section of the paper
considers experimental and observational work.
Discussion of how tests are performed,
the philosophy of broad search,
the scope of current limits and open coefficient space,
as well as some special aspects of the neutrino sector
are considered.

Though much work has been done,
a large space of opportunities remains for additional investigation.
The types of work that remain can be divided into three basic classes:
investigation at the level of the underlying theory,
model building and other investigation at the effective field-theory level,
and 
phenomenological, experimental, and observational work
performing tests within the SME framework.
Work at the underlying-theory level involves considering ways in which Lorentz violation
might arise in various candidates for the underlying theory,
such as the examples considered in section \ref{motivation},
as well as perhaps considering how Lorentz violation at that level
might be connected with low-energy physics.
At the level of effective field theory,
work remains in building additional example models,
such as the bumblebee models,
that generate the \cof\ in the SME as a result of spontaneous symmetry breaking.
Additional theoretical properties
of Lorentz violation in the SME,
such as those highlighted in section \ref{theory},
also remain to be explored.
Finally,
as noted in section \ref{currentlimits},
large sections of coefficient space remain unexplored experimentally.
In some cases,
additional phenomenology remains to identify relevant
methods of search for such coefficients in various systems.
In a number of cases,
proposals exist that await additional experimental
and observational work.

The development of the SME has provided a systematic approach
to exploring Lorentz and CPT violation in nature,
and the large amount of work that has been done
places the notion of relativity on an ever-stronger foundation.
The astoundingly sensitive explorations of Lorentz symmetry 
that have been done 
and the additional tests that are possible offer the opportunity to 
probe Planck-scale physics with existing technology.
Though no compelling evidence for Lorentz violation has been found,
the large segments of unexplored coefficient space
and the remaining theoretical questions suggest that
investigations of Lorentz symmetry will continue to play a key role
in the ongoing search for Planck-scale physics.

\end{document}